\documentclass[12pt,a4paper ]{article}

\usepackage{graphicx}

\input epsf

\newcommand{\frat}[2]{\frac{\textstyle #1}{\textstyle #2}}

\newcommand{\dmn}[2]{\mbox{$#1\!\cdot\! 10^{#2}\,$}}

\begin{document}

\begin{center}
{\Large \bf  Towards light scalar meson structure in terms of
quark and gluon degrees of freedom}\\ \vspace{0.5cm} S.V.
Molodtsov$^{1,2}$, T. Siemiarczuk$^{3}$, A.N. Sissakian$^{1}$,
A.S. Sorin$^{1}$, G.M. Zinovjev$^{4}$ \\ \vspace{0.5cm} {\small
$^1$BLTP, Joint Institute for Nuclear Research, RU-141980, Dubna, Moscow
region, RUSSIA}\\ \vspace{0.5cm} {\small $^2$Institute of
Theoretical and Experimental Physics, RU-117259, Moscow,
RUSSIA}\\ \vspace{0.5cm} {\small $^3$Soltan Institute for Nuclear
Studies, PL-00-681, Warsaw, POLAND}\\ \vspace{0.5cm} {\small
$^4$Bogolyubov Institute for Theoretical Physics, UA-03143, Kiev,
UKRAINE}
\end{center}
\vspace{0.5cm}

\begin{center}
\begin{tabular}{p{16cm}}
{\small{The origin of the lightest scalar mesons is studied in the framework
of instanton liquid model (ILM) of the QCD vacuum. The impact of vacuum
excitations on the $\sigma$-meson features is analyzed in detail. In
particular, it is noticed that the changes produced in the scalar sector may
unexpectedly become quite considerable in spite of insignificant
values of corrections to the dynamical quark masses and then the medley of
$\sigma$-meson and those excitations may reveal itself as broad resonance
states of vitally different masses.}}
\end{tabular}
\end{center}
\vspace{0.5cm}

Nowadays well-known theoretical results of existing scalar mesons,
the non-zero vacuum expectation value of which is strongly argued by
chiral symmetry breaking, find earnest experimental support. It
comes from studying the low energy $S$-wave of
$\pi$--$\pi$-scattering where the presence of solitary low mass
scalar resonance looks inevitable \cite{1} and the radiative
$\phi$-meson and heavy quarkonia decays as seen from the view point
of chiral shielding idea \cite{2}. Thus, the present situation in
the subject entirely justifies the theoretical expectation that the
physics of scalar mesons is driven by the Goldstone dynamics
\cite{3}.

However, it is still difficult to understand the properties of
scalar mesons (resonances) in terms of the QCD basic fields, in
particular, the quark and gluon origin of light $\sigma$-meson
($f_0(400$ -- $1000$ MeV)) and broad resonance $f_0(1000$ -- $1600$
MeV) \cite{4}. In a sense, the current status of the lightest
scalars could be summarized by the strange assertion that people know
where the scalars are but do not know what they are. Meanwhile, the
solution of this still pending problem can be found by
studying the QCD vacuum structure \cite{5} and searching putative
quark-gluon plasma state and in recent time the critical point in
ultra-relativistic heavy ion collisions \cite{6}. An idea to exploit
$\sigma$-meson (precisely, its coupling to photons) as a key guide
to exploring the existence and features of mixed phase of strongly
interacting matter created in nuclei-nuclei collisions at lower
energies \cite{7} requires such an insight as well. Attempts to find
an explanation of splitting two lowest scalar mesons in the
framework of rather sophisticated models mixing the quark-antiquark
states with the glueballs (see, for example \cite{8,9}) are not
fully successful and the results are ambiguous.

In the present paper we consider the origin of lowest scalar mesons in the
instanton liquid model (ILM) of the QCD vacuum \cite {10} mixing its
phonon-like excitations \cite{11} with the scalar mesons treated in a
standard way as bound quark-antiquark states. The effective Lagrangian
of phonon-like excitations is similar to the dilaton one \cite{8} and
describes the state with the quantum numbers $J^{PC} = 0^{++}$
(J = total angular momentum, P = parity, C = charge-conjugation
eigenvalue) of glueball. One specific feature of this Lagrangian is the
form of its kinetic term and rather strong interactions with quarks
(resulting in the strong mixing with the $\sigma$-meson field \cite{12}).

Avoiding the technical details of constructing this effective Lagrangian
we mention briefly here the major steps necessary for the developed approach.
In particular, it is grounded on the hypothesis the vacuum
field configurations are stabilized at the certain characteristic scale
and their action develops a well-defined minimum at the point of average
configuration size. The generating functional in ILM
$$Z=\int D[ {\cal  A}]~e^{-S( {\cal  A})}~,$$
where $S( {\cal  A})$ is the
Yang-Mills action, is supposed to be saturated by instanton
superposition
\begin{equation}
\label{1} {\cal A}^{a}_\mu(x)=\sum_{i=1}^N
A^{a}_\mu(x;\gamma_i)~.
\end{equation}
Here $A^a_{\mu}(x;\gamma_i)$ is the (anti-)instanton field in the singular
gauge
\begin{equation}
\label{2}
A^a_{\mu}(x)=\frat2g~\omega^{ab}\bar\eta_{b\mu\nu}~\frat{\rho^2}{y^2+\rho^2}
~\frat{y_\nu}{y^2}~,~~~y=x-z~,
\end{equation}
where $\gamma_i=(\rho_i,z_i,\omega_i)$ are the parameters characterizing
the $i$-th (anti-)instanton of size $\rho$ with a matrix of colour
orientation $\omega$ and with coordinates of its center position at $z$
and $g$ is the strong coupling constant. For anti-instanton the
't Hooft tensor should be substituted according to $\bar\eta \to \eta$.
We do not discuss here the mechanism of instanton ensemble stabilization
at the scale of $\bar\rho$ (see, for example \cite {13}) but focus on
extracting some phenomenological results in the context of our interest.

The form of action with peculiar minimum for the saturating configuration
equilibrated makes an existence of "oscillations" (one should keep in
mind we are working in the Euclidean space) of this configuration around
the very natural size $\bar\rho$. In principle, such a description
should be done by the corresponding Green function. However, calculating
it is not simple and eventually impracticable because the Green function
is found to be singular \cite{14}. It was noticed in Ref.\cite{15} that if
one is interested in the major terms of a generating functional exponent then
less information on interrelation between the saturating configuration
${\cal A}$ and field of "oscillations" $\frat{\partial \rho}{\partial x}$
at the characteristic scale $\bar\rho$ is necessary to calculate the
respective kinetic term of the effective Lagrangian. The instanton
ensemble excitations generated by a certain impact might naturally be
described by more general saturating configurations of the form of
Eq.({\ref 2}) if the instanton size and its colour space orientation
are varied ($\rho\to R(x,z)$, $\omega^{ab}\to\Omega^{ab}(x,z)$) and the
deformation fields could be defined by dealing with minimal action
requirement $\delta S=0$. As a result it leads to a more accurate (than
plain superposition Eq.({\ref 1})) solution and allows to determine the
interrelation looked for. In fact, it can be carried out directly
since the deformations are defined by the multipole expansion done in
the center point of instanton
\begin{eqnarray}
\label{3} R_{in}(x,z)&=&\rho+c_\mu~y_\mu+c_{\mu\nu}~y_\mu~
y_\nu+\dots~,~~~~~|y|\leq L \nonumber\\ [-.2cm]
\\[-.25cm]
R_{out}(x,z)&=&\rho+d_\mu~\frat{y_\mu}{y^2}+d_{\mu\nu}~
\frat{y_\mu}{y^2}~\frat{y_\nu}{y^2}+\dots~,~~~|y|>L~, \nonumber
\end{eqnarray}
(similar expansion should be done for instanton orientation in the
"isotopic" space $\Omega(x,z)$), here $L$ is a parameter which fixes a
sphere radius where the increasing multipole expansion with distance
increase changes its behaviour for the decreasing one because of
the requirement of deformation regularity. The coefficient $c_\mu$ in
Eq.({\ref 3}) just corresponds to the function
$\frat{\partial \rho}{\partial z_\mu}$ which we are interested in
(by the way, for the sample of deformation field Eq.({\ref 3})
it is valid $\frat{\partial R}{\partial x_\mu}\simeq-
\frat{\partial \rho}{\partial z_\mu}$ and the interrelation between two
fields is defined by the solution of Eq.({\ref 2}) with $R(x,z)$ included).

Now calculating an action of this crumpled (as was called in Ref.\cite{11})
configuration we are able to determine the kinetic term of
effective Lagrangian at the characteristic scale $\bar\rho$ (certainly,
we can not profess a higher precision in this quasi-classical
approximation as it was mentioned before) in the form
\begin{equation}
\label{4} S_{kin}=\int dx~
\frat14~G_{\mu\nu}^a(A)G_{\mu\nu}^a(A)-\beta=
\frat{\kappa}{2}~(\delta_\mu\rho)^2~,~~~\kappa=\frat{9}{10}~\beta~,
\end{equation}
here $\beta=8\pi^2/g^2$ is the single (anti-)instanton action at the
$\bar\rho$ scale which in its general form can be presented as
\begin{equation}
\label{5}
s(\rho)=\beta(\rho)+5 \ln(\Lambda\rho)-\ln \widetilde \beta^{2N_c}
+\beta \xi^2~n\bar\rho^{2}\rho^2~,
\end{equation}
with the function $\beta(\rho)=-\ln C_{N_c}-b \ln(\Lambda
\rho)~,~~ \Lambda=\Lambda_{\overline{MS}}=0.92 \Lambda_{P.V.}~,$
and the constant $C_{N_c}$ depending on the renormalization scheme
$C_{N_c}\approx\frac{4.66~\exp(-1.68 N_c)}{\pi^2
(N_c-1)!(N_c-2)!}$, $\nu=\frat{b-4}{2}$,
$b=\frat{11~N_c-2~N_f}{3}$, $N_f$ is the number of flavours, $N_c$
is the number of colours and $\beta=\beta(\bar\rho)$, $\widetilde
\beta=\beta+\ln C_{N_c}$ are the magnitudes of $\beta(\rho)$ function at
the fixed value of $\bar\rho$, $\xi$ is a constant characterizing
the repulsive power of pseudoparticles
$\xi^2=\frac{27}{4}\frac{N_c}{N_c^{2}-1} \pi^2$ and $n$ is the
instanton liquid density. Holding the terms of second order in the
small deviations from the point of action minimum $\left.\frat{d
s(\rho)}{d\rho}\right|_{\rho_c}=0$ we receive the approximate
result
\begin{equation}
\label{6}
s(\rho)\simeq s(\bar\rho)+\frat{s^{(2)}(\bar\rho)}{2}~\widetilde\varphi^2,~
\end{equation}
where $s^{(2)}(\bar\rho)\simeq\left.\frat{d^2 s(\rho)}{d\rho^2}
\right|_{\rho_c}=\frat{4\nu}{\overline{\rho^2}},$ and the scalar
field
$\widetilde\varphi=\delta\rho=\rho-\rho_c\simeq\rho-\bar\rho$ just
realizes the field of deviations from the equillibrium value
$\rho_c=\bar\rho~\left(1-\frat{1}{2\nu}\right)^{1/2}\simeq\bar\rho$.
Finally the deformation field could be described by the following
effective Lagrangian density \cite{11},\cite{15}
\begin{equation}
\label{7}
{\cal L}_\varphi=\frat{n\kappa}{2}
\left\{~\left(\frat{\partial \widetilde\varphi}{\partial z}\right)^2+
M^2_{\varphi}\widetilde\varphi^2\right\}~,
\end{equation}
with the mass gap of phonon-like excitations as
$M^2_{\varphi}=\frat{s^{(2)}(\bar\rho)}{\kappa}=
\frat{4\nu}{\kappa \overline{\rho^2}}$
(the coefficient $\kappa=0.9\beta$ generates the scale of order 1 GeV
with $\Lambda\simeq 280$ MeV).

Then the quark determinant ${\cal Z}_\psi$ for the stochastic ensemble of
pseudo-particles looks like
$${\cal Z}_\psi~\simeq~\int D\psi^\dagger D\psi~\langle\langle~
e^{S(\psi,\psi^\dagger,A)}~\rangle\rangle_A~,
$$
where $S(\psi,\psi^\dagger,A)$ is the action of QCD with massless quarks
which should describe the spontaneous breaking of chiral symmetry and
generation of dynamical quark mass of order $300$ MeV \cite{16}.
Eventually the quark generating functional (after absorbing the variations of
(anti-)instanton average size) takes the following form (for $N_f=2$)
of the integral over saddle point parameters \cite{12}
\begin{eqnarray}
\label{8}
&&{\cal Z}_\psi=\int d\lambda~D M~D\psi^\dagger D\psi~e^{-{\cal L}_{\psi,
\widetilde\varphi}}~,\nonumber\\
&&{\cal L}_{\psi, \widetilde\varphi}=-N\ln\frat{N}{\lambda V}+ 2 \int dx
~(N_f-1)~\lambda^{-\frac{1}{N_f-1}}~
(Det M)^{\frac{1}{N_f-1}}-\\
&&-\int \frat{dkdl}{(2\pi)^8}~
\psi_{f}^\dagger(k)\left[(2\pi)^4\delta(k-l)\left(-\hat k+
i~ m_{fg}~v(k,k)\right)+i~
m_{fg}~u(k,l)~\widetilde\varphi(k-l)\right]\psi_g(l)~,\nonumber
\end{eqnarray}
where $N$ is the total number of pseudo-particles in the volume $V$
($n=N/V$), $M$ is the $N_f\times N_f$-matrix of scalar boson fields in the
flavour space and $f,g$ are the flavour indecies. The vertex functions
$v(k,k)$ and $u(k,l)$ are defined by the zero-modes (those are the
solutions of the Dirac equation in the field of (anti-)instanton of zero
energy) and have the following forms
$$v(k,k)=G^2(k),~~~
~u(k,l)=\frat{d~ v(k,l)}{d\rho}=G(k)G'(l)+G'(k)G(l)~,$$
moreover $G(k,\rho)=2\pi\rho F(k\rho/2)$,
$G'(k,\rho)=\frat{d G(k,\rho)}{d\rho}$,
where
$$
F(x)=2x~[I_0(x)K_1(x)-I_1(x)K_0(x)]-2~I_1(x)K_1(x)~,
$$
and $I_i,~K_i~(i=0,1)$ are the modified Bessel functions.

After executing the bosonisation procedure the Lagrangian density takes
the following form
\begin{equation}
\label{9}
{\cal L}={\cal L}_{\widetilde\varphi} + {\cal L}_{\psi, \widetilde\varphi}~,
\end{equation}
which describes the quarks and phonon-like excitations of instanton
liquid. The Lagrangian parameters are fixed by saddle point available as
$$\frat{\partial {\cal L}}{\partial \lambda}=0~~,~~~~\frat{\partial {\cal
L}}{\partial M}=0~.$$
In the particular case of negligible impact of field $\widetilde\varphi$,
saddle point values of parameters are fixed by the requirements
$$\lambda^{-\frac{1}{N_f-1}}~\prod_{i=1}^{N_f}
m_{i}^{\frac{1}{N_f-1}}=\frat{N}{2V}~,$$
\begin{equation}
\label{10}
4 N_c \int
\frat{dk}{(2\pi)^4}~\frat{m_f^{2}~v^2(k)}{k^2+m_f^{2}~v^2(k)}=\frat{N}{V}~~,
\end{equation}
here $v(k)=v(k,k)$.

As it was mentioned above the Lagrangian density ({\ref 9}) is quite
similar to the dilaton one. However, there are, at least, two substantial
distinctions. First, it is the presence of factor $n \kappa$ in the
kinetic term. This factor is analogous to the factor $F^2_{\pi}$ of chiral
$\pi$-meson Lagrangian density
${\cal L}_\pi=\frat{F^2_{\pi}}{2}~\partial_\mu \widetilde\pi \partial_\mu
\widetilde\pi$,
and sets up the scale of phonon-like excitations. Average instanton size $\bar\rho$
($\widetilde\varphi=\bar\rho\varphi$)
is quite relevant dimensional unit for phonon-like field and, hence, the
"correct" dimension for the factor in kinetic term could be
$$F^2_{\varphi}=n \kappa \bar\rho^2~.$$
The second distinction concerns the interrelation of phonon-like field
and quarks which does not disappear in the chiral limit and is defined
by the vertex function $u(k,l)$.

Then a nontrivial solution of saddle point equation Eq.({\ref 10}) fixes
the dynamical quark mass $M(k)=m_f v(k)$ together with the quark condensate
$$-i\langle\psi^\dagger\psi\rangle=-4N_c~\int \frat{dk}{(2\pi)^4}\frat{M(k)}
{k^2+M^2(k)}~.$$
The excitations of quark condensate with the respective quantum numbers
are seen as $\sigma$- and $\pi$-mesons
(in this paper we are dealing with $N_f=2$ only) and
$$M=m~(1+\widetilde\sigma)~e^{i \widetilde\pi^a \tau^a}~,$$
where $\tau^a$ are the Pauli matrices. The corresponding correlation
functions $R_\sigma$ and $R_\pi$ are given, as known, in the form \cite{16}
\begin{eqnarray}
\label{11}
&&R_\pi(p)= 2 N_f N_c \int
\frat{dk_1}{(2\pi)^4}~\frat{M_1^{2}}{k_{1}^2+M_1^{2}}-
2 N_f N_c \int \frat{dk_1}{(2\pi)^4}~\frat{((k_1 k_2)+M_1 M_2)~M_1 M_2}
{(k_{1}^2+M_1^{2})(k_{2}^2+M_2^{2})}~~,\\
\label{12}
&&R_\sigma(p)= n\bar\rho^4-
2 N_f N_c \int \frat{dk_1}{(2\pi)^4}~\frat{((k_1 k_2)-M_1 M_2)~M_1 M_2}
{(k_{1}^2+M_1^{2})(k_{2}^2+M_2^{2})}~~,
\end{eqnarray}
with $k_2=k_1+p$, $M_1=M(k_1)$, $M_2=M(k_2)$.
Expanding the integrals around small values of momentum $p$ as
$$R_\pi(p)=\beta_\pi~p^2~,~~~R_\sigma(p)=\alpha_\sigma+\beta_\sigma~p^2~,$$
we are able to calculate the following decay constants
\begin{eqnarray}
&&\beta_\pi=\frat{F_{\pi}^2}{2}=N_c N_f~ \int
\frat{dk}{(2\pi)^4}~\frat{M^2(k)-\frat{k}{2}~M'(k) M(k)
+\frat{k^2}{4}~(M'(k))^2}{(k^2+M^2(k))^2}~,~\nonumber\\
&&\beta_\sigma=\frat{F_{\sigma}^2}{2}=\beta_\pi+4 N_c N_f~ \int
\frat{dk}{(2\pi)^4}~\left[\frat{2
M^2(k)}{(k^2+M^2(k))^2}~\Delta_1-
\frat{M^4(k)}{(k^2+M^2(k))^4}~\Delta_2\right]~,~\nonumber\\
&&\Delta_1=\frat{1}{16}~\left(\frat{3}{k}
~M(k)M'(k)-M'(k)M'(k)+M(k)M''(k)\right)~,\nonumber\\
&&\Delta_2=\frat{k^2+M^2(k)}{2}~\left(1+\frat{M(k)M'(k)}{k}\right)-
\frat{k^2}{4}~\left(1+\frat{M(k)M'(k)}{k}\right)^2+\nonumber\\
&&+\frat{k^2+M^2(k)}{8}~\left(M'(k)M'(k)+M(k)M''(k)-\frat{M(k)M'(k)}{k}\right)~,\nonumber
\end{eqnarray}
(here the prime denotes the derivative calculation) and the $\sigma$-meson
mass as well
\begin{eqnarray}
&&M_\sigma^{2}=\frat{\alpha_\sigma}{\beta_\sigma}~,\nonumber\\
&&\alpha_\sigma=n\bar\rho^4-2 N_f N_c \int
\frat{dk}{(2\pi)^4}~\frat{(k^2-M^2(k))~M^2(k)}
{(k^2+M^{2}(k))^2}~.\nonumber
\end{eqnarray}
The width $\Gamma_\sigma$ of $\sigma$-meson decay into $\pi\pi$ is defined
by the set of graphs $\alpha$---$\delta$ (see Fig.1) for the anharmonic
term $\sim\widetilde\sigma\widetilde\pi^2$ of effective Lagrangian
$${\cal L}_{int}=\int \frat{dm dl}{(2\pi)^8}\int \frat{dk}{(2\pi)^4}~
W_\sigma(k,m,l)~\widetilde\sigma(m)~\widetilde\pi^a(l)~\widetilde\pi^a(m-l)~
,$$
\begin{eqnarray}
\label{13}
&&\frat{W_\sigma(k,m,l)}{4 N_c}=-\frat{M^2(k)}{k^2+M^2(k)}-\nonumber\\
&&-\frat{k(k-m)-M(k)M(k-m)}{(k^2+M^2(k))((k-m)^2+M^2(k-m))}~M(k)M(k-m)+\nonumber\\
&&+2~\frat{k(k-l)+M(k)M(k-l)}{(k^2+M^2(k))((k-l)^2+M^2(k-l))}~M(k)M(k-l)+
\nonumber\\
&&+2~\frat{k(k-m) M(k-l)-k(k-l) M(k-m)-((k-m)(k-l) M(k)-M(k) M(k-m) M(k-l)}
{(k^2+M^2(k))((k-m)^2+M^2(k-m))((k-l)^2+M^2(k-l))}\times\nonumber\\
&&\times M(k)M(k-m)M(k-l)~. \nonumber
\end{eqnarray}
(four terms of this expression are in explicit correspondence with the four
graphs of Fig.1).

\begin{tabular}{ll}
\begin{picture}(100,100)(10,20)
\put(15,30){\line(2,0){50}}
\put(15,28){\line(2,0){50}}
\put(65,28){\line(2,-1){10}}
\put(82.5,19.5){\line(2,-1){10}}
\put(65,30){\line(5,-1){10}}
\put(82.5,26.5){\line(5,-1){10}}
\put(65,50){\circle{100}}
\put(65,0){$\alpha$}
\end{picture}
&
\hspace{3cm}
\begin{picture}(100,100)(10,20)
\put(-30,50){\circle{100}}
\put(-80,49){\line(2,0){30}}
\put(-80,51){\line(2,0){30}}
\put(-10,49){\line(2,-1){10}}
\put(10,39){\line(2,-1){10}}
\put(-10,51){\line(2,1){10}}
\put(10,61){\line(2,1){10}}
\put(-30,0){$\beta$}
\put(95,50){\circle{100}}
\put(45,49){\line(2,0){30}}
\put(45,51){\line(2,0){30}}
\put(115,50){\line(2,0){10}}
\put(135,50){\line(2,0){10}}
\put(45,66){\line(2,-1){10}}
\put(65,56){\line(2,-1){10}}
\put(95,0){$\gamma$}
\end{picture}
\end{tabular}

\begin{tabular}{ll}
\begin{picture}(100,100)(10,20)
\put(15,51){\line(2,0){50}}
\put(15,49){\line(2,0){50}}
\put(65,49){\line(2,-1){50}}
\put(65,51){\line(2,1){50}}
\put(115,24){\line(0,2){52}}
\put(115,76.5){\line(2,0){10}}
\put(135,76.5){\line(2,0){10}}
\put(115,24){\line(2,0){10}}
\put(135,24){\line(2,0){10}}
\put(65,0){$\delta$}
\end{picture}
&
\hspace{3cm}
\begin{picture}(100,100)(10,20)
\put(55,30){\line(2,0){50}}
\put(55,28){\line(2,0){50}}
\put(105,29){\line(2,0){5}}
\put(115,29){\line(2,0){5}}
\put(125,29){\line(2,0){5}}
\put(135,29){\line(2,0){5}}
\put(145,29){\line(2,0){5}}
\put(105,50){\circle{100}}
\put(50,0){
}
\end{picture}
\end{tabular}
\begin{figure*}[!tbh]
 \caption{The single line corresponds to the quark contribution, the double one
corresponds to $\sigma$-meson, the short dashed line shows the phonon-like
field and the long dashed one shows $\pi$-meson. }
  \label{fig1}
\end{figure*}

Then for the decay width of $\sigma$-meson in its rest frame
we find that after integrating the function
$W_\sigma(k,m,l)/F_{\sigma}/F^2_{\pi}$
over the loop at fixed momenta of $\pi$-meson \cite{17} we have
$$\Gamma_\sigma=\frat{3}{8\pi}\frat{W_\sigma^{2}}{M_\sigma}~\left(1-\frat{4
M_\pi^{2}}{M_\sigma^{2}}\right)^{1/2}~,$$
here $W_\sigma$ denotes the integral over anharmonic contribution of type
$W_\sigma~\sigma\pi^2$. At zero outgoing momenta the anharmonic
contribution disappears $W_\sigma(k,0,0)=0$ because in the chiral limit $\pi$-meson fields
may appear in the combination with the derivative term $(\partial_\mu\pi)^2$
only. As known, this fact, in some extent, results in the small $\sigma$-meson width.

In the meantime, if the phonon-like excitations are not taken into
account $\pi$ and $\sigma$-mesons are defined by the following fundamental
characteristics
\begin{eqnarray}
&&F_\pi\simeq 121~ {\mbox{MeV}}~,~~F_\sigma\simeq
96~{\mbox{MeV}}~,\nonumber\\
&&M_\sigma\simeq 550~{\mbox{MeV}}~,~~\Gamma_\sigma\simeq
2.8~{\mbox{MeV}}~,\nonumber
\end{eqnarray}
and $\pi$-meson mass can be extracted from Gell-Mann-Oakes-Renner (GMOR)
relation.The saddle-point parameter $m$ taken as dimensionless quantity is
equal to $m=\dmn{4.817}{-3}$. Then we have for the instanton liquid density
$n/\Lambda^4=1.03$, for the average pseudo-particle size
$\bar\rho\Lambda=0.28$ and eventually for single instanton action at this
characteristic scale $\beta(\bar\rho)=19$. The dynamical quark mass and
quark condensate read as
$$M\simeq 385~{\mbox{MeV}}~,~~-i\langle\psi^\dagger\psi\rangle\simeq
 -(381)^3~{\mbox{MeV}}^3~,$$
at $\Lambda =280$ MeV. Surprisingly, the small width of $\sigma$-meson is a
result of interference of all terms contributing. In particular, in
dimensionless units the corresponding contributions are as follows
$$\alpha=\dmn{-1.199}{-2},~~\beta=\dmn{-7.137}{-3},~~\gamma=\dmn{2.066}{-2},
~~\delta=\dmn{-1.695}{-3},~~
\alpha+\beta+\gamma+\delta=\dmn{-1.648}{-4}~.$$

Actually, many quantitative estimates are based on the calculations
of $\delta$ graph only and it results in the magnitude of width
$\Gamma_\sigma$ which is declared to be a few hundred MeV. Assuredly, there
is an unpleasant drawback of this calculation. It comes from the
opposite signs of contributions of $\alpha$ and $\gamma$ diagrams because the
obvious compensation produces rather small value which is poorly
controlled by one-loop calculations. Unfortunately, for the time being two-loop
calculations are hardly realized technically.

In order to include the changes coming from phonon-like fields entering
the game, the perturbation scheme was proposed in Ref.\cite{12} in the
tadpole approximation. Actually, it becomes possible, because of
the quark condensate presence in the following sum
$$\psi^\dagger\psi~\varphi=\langle\psi^\dagger\psi\rangle~\varphi+
(\psi^\dagger\psi-\langle\psi^\dagger\psi\rangle)~\varphi~,$$
to keep the first term only in the respective inserts into quark Green
functions and vertices. Then the dynamical quark mass is defined not only
by the vertex $v$ but the tadpole contribution with vertex $u$ as
well (see Eq.(\ref{8})). Amazingly, the dependence of diagram contributions on
the coefficient $\kappa$ is cancelled in this approximation. Generally,
there is hardly any impact on the instanton liquid (it is
negligible), average instanton size and parameter $\beta(\bar\rho)$ do
not change but instanton liquid density slightly increases
$n/\Lambda^4=1.11$.
The saddle point parameter is getting smaller $m=\dmn{3.481}{-3}$ which
results in the dynamical quark mass and quark condensate decreasing (see
Fig. 2).
$$M\simeq 324~{\mbox{MeV}}~,~~-i\langle\psi^\dagger\psi\rangle\simeq
 -(343)^3~{\mbox{MeV}}^3~.$$

\begin{figure*}[!tbh]
\begin{center}
\includegraphics[width=0.5\textwidth]{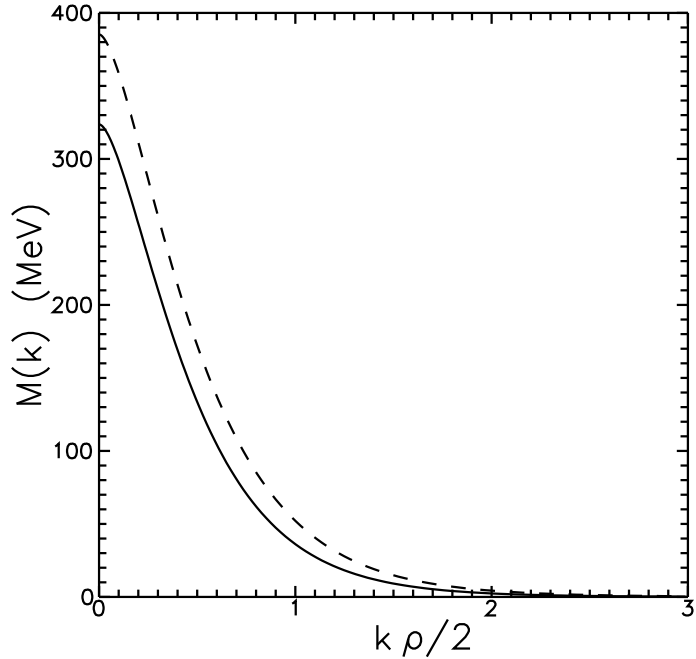}
\end{center}
 \vspace{-7mm}
\caption{The dynamical quark mass as a function of momentum. The dashed
line shows that behaviour when the phonon-like fields are not taken into
account.}
 \label{mass}
\end{figure*}

However this quite noticeable decrease of quark dynamical mass (and quark
condensate as well) should not be taken very seriously because the scale
of $\Lambda$ in the ILM is not strictly fixed and it is normalized
by calculating the observables. Thus, if one considers, for example,
$\pi$-meson there is only the possibility to fit parameters (for both
with phonon-like excitations of instanton liquid and without them) in such
a way to get the satisfactory value of constant $F_\pi$, the GMOR
relation and etc. Then characteristic parameters of field scales in the
tadpole approximation are the following
$$F_\pi\simeq 106~ {\mbox{MeV}}~,~~F_\sigma\simeq
86~{\mbox{MeV}}~,~~F_\varphi\simeq 337~{\mbox{MeV}}~.$$
Peculiarly, the field $\varphi$ develops remarkably larger scale
comparing to $\pi$- and $\sigma$-mesons.

The width $\Gamma_\varphi$ of phonon-like field $\varphi$ decay into
$\pi\pi$-mode is defined by the diagrams similar to $\alpha$ and
$\gamma$ but with changing $\sigma$-meson line for the line of phonon-like field.
Then the corresponding anharmonic term of effective Lagrangian density
$\widetilde\varphi\widetilde\pi^2$ takes the form as
$${\cal L}_{int}=\int \frat{dm dl}{(2\pi)^8}\int \frat{dk}{(2\pi)^4}~
W_\varphi(k,m,l)~\widetilde\varphi(m)~\widetilde\pi^a(l)~\widetilde\pi^a(m-l )~,$$
\begin{eqnarray}
\label{14} &&\frat{W_\varphi(k,m,l)}{4 N_c}=-\frat{M(k)~\Delta
M(k)}{k^2+M^2(k)}+\nonumber\\
&&+2\frat{k(k-l)+M(k)~M(k-l)}{[k^2+M^2(k)][(k-l)^2+M^2(k-l)]}~M(k,k-l)~\Delta
M(k-l,k)~, \nonumber
\end{eqnarray}
two terms of the sum here correspond to the diagrams mentioned and
$\Delta M(k)=m_f~u(k,k)$. Finally, we have
$$\Gamma_\varphi=\frat{3}{8\pi}\frat{W_\varphi^{2}}{M_\varphi}~\left(1-\frat
{4 M_\pi^{2}}{M_\varphi^{2}}\right)^{1/2}~,$$ where $W_\varphi$
looks similar to $W_\sigma$ but with changing the factor in
denominator providing the dimensionless result for $F_{\varphi}
F^2_{\pi}$). The mass of phonon-like field and its width are
$$M_\varphi\simeq 827~{\mbox{MeV}}~,~~\Gamma_\varphi\simeq
340~{\mbox{MeV}}~,$$ but these characteristics for $\sigma$-meson
become as $$M_\sigma\simeq
1.004~{\mbox{MeV}}~,~~\Gamma_\sigma\simeq 270~{\mbox{MeV}}~,$$
(for quark loop without the tadpole contribution the width of
phonon-like excitation is estimated to be $\Gamma_\varphi\simeq
780~{\mbox{MeV}}$). The partial contributions of diagrams to the
$\sigma$-meson width read as
$$\alpha=\dmn{-7.508}{-3},~\beta=\dmn{-2.205}{-3},~\gamma=\dmn{8.567}{-3},~\
\delta=\dmn{-2.999}{-4},~~
\alpha+\beta+\gamma+\delta=\dmn{-1.447}{-3}.$$ It is instructive
to compare the magnitudes of $\alpha_\sigma$ and $\beta_\sigma$
integrals contributing dominantly to the $\sigma$-meson mass
(upper line corresponds to the contribution without phonon-like
field)
$$\alpha_\sigma=\dmn{1.333}{-3},~\beta_\sigma=\dmn{1.812}{-2},~
\beta_\pi=\dmn{2.872}{-2}~,$$
$$\alpha_\sigma=\dmn{3.508}{-3},~\beta_\sigma=\dmn{1.427}{-2},~
\beta_\pi=\dmn{2.202}{-2}~,$$ (for all that the packing fraction
parameters are $n\bar\rho^4=\dmn{5.997}{-3}$ for the upper line
and $n\bar\rho^4=\dmn{6.494}{-3}$ for the lower one). As seen, the
coefficient $\alpha_\sigma$ gets the strongest impact.

Now let us turn to the problem of mixing two scalar fields if their
Lagrangian density tolerating the decay into two $\pi$-mesons mode has
the following form
\begin{eqnarray}
\label{15}
&&-\widetilde{\cal L}=-\frat{F^2_{\varphi}}{2}~(k^2+M_\varphi^{2})~\widetilde\varphi^2
 -\frat{F^2_{\sigma}}{2}~(k^2+M_\sigma^{2})~\widetilde\sigma^2-\nonumber\\
[-.2cm]
\\ [-.25cm]
&&-\frat{F^2_{\pi}}{2}~(k^2+M_\pi^{2})~\widetilde\pi^2+\widetilde\Delta~
\widetilde\varphi \widetilde\sigma+
W_\varphi~\widetilde\varphi\widetilde\pi^2+
W_\sigma~\widetilde\sigma\widetilde\pi^2~.\nonumber
\end{eqnarray}
Here the parameter regulating the component mixing $\widetilde\Delta$ is
given by the tadpole graph of Fig.1 (and the sign of mixing term coincides
with the sign of quark condensate) and could be presented as
$$ \widetilde\Delta=4 N_c N_f~\int
\frat{dk}{(2\pi)^4}~\frat{m_f~u(k)~M(k)}{k^2+M^{2}}~.$$
As it was mentioned above, unlike the dilaton Lagrangian \cite{8} in this
scheme the field mixing survives in the chiral limit and for the instanton
liquid parameters used above we have
$\Delta=\frat{\widetilde\Delta}{F_\sigma F_\varphi}= (483~{\mbox{MeV}})^2$.
The masses of diagonal fields are given by
\begin{equation}
\label{16}
M_{\pm}^2=\frat{M_\varphi^{2}+M_\sigma^{2}}{2}\pm
\frat{((M_\varphi^{2}+M_\sigma^{2})^2+4\Delta^2)^{1/2}}{2}~.
\end{equation}
Then the minimal mass of light scalar component is limited by constraint
$M_-=2 M_\pi$, otherwise it could be stable in strongly interacting mode,
and the upper limit of the $\Delta$ magnitude responsible for mixing
looks like
$$\Delta_{max}^2\leq (M_\varphi^{2}-4M_\pi^{2})(M_\sigma^{2}-4M_\pi^{2})~,$$
what for existing parameters gives $\Delta_{max}=(868~{\mbox{MeV}})^2$.
Comparing to the standard $\Delta$ value one may conclude the mixing effect
is quite significant.

Passing through the standard procedure with $\theta$ as a mixing angle
\begin{eqnarray}
&&\varphi=\cos \theta ~\varphi'+\sin \theta~ \sigma'~,\nonumber\\
&&\sigma=-\sin \theta~ \varphi'+\cos \theta ~\sigma'~,\nonumber
\end{eqnarray}
we receive for corresponding widths of $\varphi'$ and $\sigma'$
the following results $$W_{\varphi'}=W_\varphi \cos \theta -
W_\sigma \sin \theta~,~~ W_{\sigma'}=W_\varphi \sin \theta +
W_\sigma \cos \theta~,$$ where the value of mixing angle could be
obtained from $$tg
(2\theta)=\frat{2~\Delta}{M_\sigma^{2}-M_\varphi^{2}}~.$$ It
becomes clear from Eq.(\ref{14}) the mass of lighter components is
getting smaller whereas the heavier one gains more mass.
Remembering the phonon-like excitation was lighter before mixing
we find out for $\varphi'$ the following characteristics
$$M_{\varphi'}\simeq 749~{\mbox{MeV}}~,~~\Gamma_{\varphi'}\simeq
265~{\mbox{MeV}}~.$$ The heavier component $\sigma'$ becomes as
$$M_{\sigma'}\simeq 1.063~{\mbox{MeV}}~,~~\Gamma_{\sigma'}\simeq
175~{\mbox{MeV}}~.$$ The mixing angle is $\theta=27.5$ degrees.
Unfortunately, the precision of calculating the particle decay
widths is not high and allows to rely on the order of magnitude
obtained only. The reason is tightly linked to the contributions
of the $\alpha$ and $\gamma$ graphs which are considerably larger
than others. Calculating the $\alpha$ and $\gamma$ contributions
within 10 \% precision gains a much poorer precision for the total
result. In general it is rooted in that fact the estimates are
obtained at the limit (sometimes beyond it in other approaches) of
applicability of expansions used.

Summarizing, we would like to underline the approach proposed here sheds more
light on the possible origin of scalar mesons. Assuredly, many other models have
expressed the similar intends and succeeded to different extents. However, the
calculations carried out in this paper demonstrate a quite robust effect of
phonon-like excitations of instanton liquid in the scalar sector and capacity
of such a model to successfully absorb new results and hints of running experiments.
It concerns, first of all, the meson state splitting and its strong coupling scale.
For example, both quark-antiquark and glueball $0^{++}$ operators mixed to create
the scalar state in lattice QCD calculations give the mass of the lightest meson
essentially suppressed with respect to the mass of corresponding ($0^{++}$) glueball
and demonstrate strong lattice spacing dependence \cite{18}. Besides, the model
developed in this paper can be easily adapted for studying the influence of hot
and dense medium which appears in ultrarelativistic heavy ion collisions and what
could be fairly crucial step in accomplishing the objective.

This research was supported by the Grants INTAS-04-84-398, NATO
PDD(CP)-NUKR980668, RFBR 05-02-17695 and by the special program of the Ministry
of Education and Science of the Russian Federation (grant RNP.2.1.1.5409).
We most gratefully acknowledge the discussions with A.E. Dorokhov, P. Giubellino,
E.A. Kuraev, A.E. Radzhabov,  M.K. Volkov and V.L. Yudichev.


\end{document}